# A Directly Coupled Superconducting Quantum Interference Device Magnetometer Fabricated in Magnesium Diboride by Focused Ion Beam

Gavin Burnell, Dae-Joon Kang, and David A Ansell

Dept. Materials Science and IRC in Superconductivity,

University of Cambridge, Pembroke St., Cambridge CB2 3QZ, UK

H.-N. Lee and S.-H. Moon

LG Electronics Institute of Technology

Seoul 137-724, Korea

Edward J Tarte and Mark G Blamire

Dept. Materials Science and IRC in Superconductivity,

University of Cambridge,

Pembroke St., Cambridge, CB2 3QZ

We report the fabrication of a directly coupled superconducting quantum interference device (SQUID) magnetometer in $MgB_2$ using a focused ion beam (FIB) to create Josephson junctions in a 70 nm thick film of $MgB_2$. The SQUID shows a voltage modulation ($\Delta V$) of 175 $\mu V$ at a temperature of 10 K and operates over a temperature range from 10 K to 24 K. We find excellent agreement between the measured maximum transfer functions and those predicted by theory. We have measured the magnetic flux noise at 20 K to be as low as 14 $\mu\Phi_0 Hz^{-1/2}$.

Since the discovery that $MgB_2$ was a superconductor with a critical temperature ($T_C$) of 39 K[1], there has been, in addition to fundamental studies, considerable interest in developing the material for practical applications. Tunneling studies[2-5] indicate that $MgB_2$ is reasonably isotropic with an s-wave character – albeit with two gaps arising from multi-band superconductivity[6,7]– implying that superconducting devices based on $MgB_2$ could operate up to 30 K. There have been a number of reports on the growth of thin films either by deposition of a precursor followed by annealing in an Mg rich environment[8-12] or by direct deposition of $MgB_2$[13-15], however there have been few reports of all-$MgB_2$ thin film junctions and devices[16,17]. $MgB_2$ SQUIDs based on thin film nanobridges[17] and point contact junctions[18] have been reported, in the former case the relatively small voltage modulation limits the utility of the device, and the latter whilst showing impressive performance has a physical configuration that is unsuitable for many applications.

We have used our previously reported[16] technique for fabricating $MgB_2$ superconductor-normal metal superconductor (SNS) Josephson junctions using localized ion damage caused by a FIB to create a single-layer thin film SQUID with a directly coupled pick-up loop of outer dimension 2 mm on which we report in this letter.

Our thin film deposition and fabrication technique have been fully described elsewhere[10,16]. The device design used in this work was similar to that used for previous work on $YBa_2Cu_3O_{7-\delta}$ grain boundary SQUIDs[19]. The SQUIDs consisted of a square pick-up coil of outer dimension 2mm and width 750 µm. The SQUID loop itself had a slit-type geometry, with a line width of 4 µm and a inner dimension of 5 µm. The length of the SQUID loop was such as to achieve a 100 pH inductance (geometric plus kinetic). This particular SQUID design was intended to have a resistor connected across the loop close to the junctions in a configuration as suggested by Enpuku[20], in this work we severed this link by milling a 250 nm break using the FIB.

In order to fabricate the junctions, the device was wire bonded to a custom holder and transferred to our focused ion beam system (Philips-FEI Inc. FIB 200). The junctions were defined by writing cuts of width 50 nm across the width of the tracks in the SQUID loop using a 4 pA, 30 kV Ga ion beam. The depth of the cut was controlled by monitoring the change in resistance of the device during the FIB milling process[21] and the milling was stopped when a predetermined resistance change was achieved. We found that a resistance-change track-width product of 30-40 Ω µm gave junctions with critical-current ($I_C$) per unit track width of order 100 µA µm$^{-1}$ at 10 K. By comparing the time taken to achieve this resistance change with the time to completely sever a track and assuming a constant milling rate, we were able to deduce that 15-20 nm of $MgB_2$ remained at the bottom of the cut trench as in our previous work[16]. In Fig. 1 we show a FIB image of the SQUID loop region and inset a schematic of the complete SQUID including pickup coil.

After fabrication, the current-voltage (*I-V*) characteristics and voltage-flux (*V-Φ*) curves were measured between 10 and 24 K. Below 10 K the *I-V* characteristics were hysteretic and thus the V-Φ curves, whilst showing peak-to-peak amplitudes of several hundred μV, were unstable. In Fig 2 we show a series of V-Φ curves for the SQUID at 10 K at various current bias points and inset the I-V characteristics between 10 and 16 K. The largest ΔV was 175 μV for a current bias of 464 μA or 1.02 $I_C$ and temperature of 10 K. The V-Φ curves are offset with respect to the applied field such that the minimum does not coincide with zero applied field as the SQUID design (Fig. 1) is such that the SQUID bias current contributes to the field applied to the SQUID. The slight asymmetry in the curves would suggest that the two junctions have different $I_C$[22]. The I-V characteristic shows a 'bump' at 0.75 mV, which we attribute to a resonance feature.

From the V-Φ curves we can extract the transfer function $V_\Phi = \left|\partial V / \partial \Phi\right|$, the amplitude of which we show in Fig. 3 in addition to the magnitude of ΔV and $\beta_L = 2LI_0/\Phi_0$ for temperatures between 10 and 26 K (where L is the SQUID inductance, $I_0$ is the ciritical current for one junction so that $2I_0=I_C$ the critical current for the SQUID and $\Phi_0$ the flux quantum). In order to compare the maximum $V_\Phi$ with that predicted from fits to numerical models of SQUIDs based on the resistively shunted junction (RSJ) model, it is useful to consider a normalized transfer function amplitude:

$$v_\Phi = 3V_\Phi / I_C R_d \quad (1)$$

(where $I_C R_d$ is the product of the critical current and dynamic resistance at the current bias point for the SQUID) as a function of $\Gamma\beta_L$ (where the noise parameter $\Gamma = 4\pi k_B T / 2I_0\Phi_0$ ). It should be noted that conventionally the transfer function amplitude is normalized to $I_C R_N$ where $R_N$ is the normal state resistance for the SQUID. When a SQUID is biased in the resistive regime, the measured voltage changes depend on the changes in the dc and time averaged ac Josephson currents and also the dynamic resistance. In the RSJ model, the dynamic resistance scales linearly with $R_N$ so it is appropriate use $R_N$ as the normalization constant. In our devices there is a resonance feature due to the capacitive and inductive impedances associated with the SQUID slit and severed resistive short. This resonance feature manifests as a 'bump' in the I-V characteristic at biases slightly larger than the optimum SQUID bias point. The I-V characteristics shown in the inset to Fig 2 clearly illustrate the effect of this resonance feature in reducing the dynamic resistance just above the $I_C$ and that this reduction is more significant at higher temperatures. Therefore we normalize to the local dynamic resistance. In order to compare with the fits based on RSJ models normalized to $R_N$, however, we also need to include the ratio of $R_d$ to $R_N$ in the RSJ model. For an RSJ-like junction biased at 1.05 $I_0$ the dynamic resistance is 3 times the $R_N$, thus we normalize to $I_C R_d/3$. The inset to Fig 3 shows $v_\Phi$ against $\Gamma\beta_L$ and also the fit found by Enpuku[23]. There is excellent agreement between our experimental results and the numerical simulations, however it should be noted that whilst the approximation used by Enpuku:

$$v_\Phi = \frac{4}{1+\beta_L} e^{-2.75\Gamma\beta_L} \quad (2)$$

agrees with the numerical results for 0.1<$\Gamma\beta_L$<0.4, as $\Gamma\beta_L$ tends to 0.04, it increasingly underestimates the numerically modeled transfer function (by up to 25% for $\Gamma\beta_L$ =0.04) [24], so that the agreement with our data indicates that there is another factor reducing the transfer function at lower $\Gamma\beta_L$.

To measure the noise performance of our SQUID, we operated it in a flux-locked loop at a temperature of 20 K in a probe equipped with three layers of high permeability µ-metal shielding. In Fig. 4 we show the noise spectrum recorded from the device. For frequencies below 150 Hz, the device shows a 1/f frequency dependent noise. Above this frequency the noise is only very weakly dependent on frequency falling from $S_\Phi^{1/2}$= 20 µ$\Phi_0$Hz$^{-1/2}$ at 1 kHz to 14 µ$\Phi_0$Hz$^{-1/2}$ at 10 kHz; the peak above 20 kHz and subsequent drop off are artifacts of the electronics. Comparing the white noise with theory[24], taking

$$S_\Phi^{1/2} = \left(\frac{16 k_B T R}{V_\Phi^2}\right)^{1/2} \qquad (3)$$

and substituting in the dynamic resistance at the SQUID bias and the maximum measured $V_\Phi$, we find $S_\Phi^{1/2}$=1.7 µ$\Phi_0$Hz$^{-1/2}$. The extra order of magnitude noise measured is consistent with the only other noise results for a MgB$_2$ SQUIDs in the literature[18] and is likely to be a result of similar causes to those cited in that work.

In summary, we have fabricated a directly coupled SQUID magnetometer in thin film MgB$_2$ using a focused ion beam to create two SNS junctions. The resultant SQUID shows voltage modulations up to 175 µV at 10K. We have been able to show that the maximum transfer function is consistent with results of numerical simulations given in the literature if one uses the dynamic resistance at the SQUID bias point when normalizing the experimental data. Finally, we have measured the noise of the SQUID operating at 20K and find that it is consistent with that reported for SQUIDs made by push junctions and comparable to high temperature superconductor (HTS) SQUIDs; YBa$_2$Cu$_3$O$_{7-\delta}$ SQUIDs made with grain boundary junctions and using the same design show a noise of 9.1 µ$\Phi_0$Hz$^{-1/2}$ at 77K[19]. The results presented in this work are very encouraging for the development of MgB$_2$ as a material for superconducting electronics and devices operating at 20 to 30 K.

This work was funded by the UK Engineering and Physical Sciences, and Particle Physics and Astronomy Research Councils and the Korean Ministry of Science and Technology under the National Research Laboratory Project.

Figure Captions

**Focused Ion Beam microscope image of a SQUID showing the slit-geometry loop, asymmetrically attached pickup coil and resistive short link. Inset, schematic of complete design showing the 2mm outer dimension pickup coil.**

**Voltage modulation with applied magnetic flux at 10 K for current biases of 490 (top line), 480, 470, 468, 466, 464, 462, 460, 450, 440 (bottom curve) μA. Inset, current-voltage characteristics for 10, 12, 14 and 16 K (order indicated by arrow).**

**Maximum voltage modulation (triangles, dotted line), maximum transfer function (diamonds, solid line) and $\beta_L$ (circles, dashed line and right hand scale) between 10 and 26 K. Lines are a guide for the eye only. Inset, normalized maximum transfer function with $\Gamma\beta_L$ (squares - experimental data, line is the approximation to numerical simulations by Enpuku).**

**Flux noise spectra of SQUID measured at 20 K using a flux-locked loop scheme.**

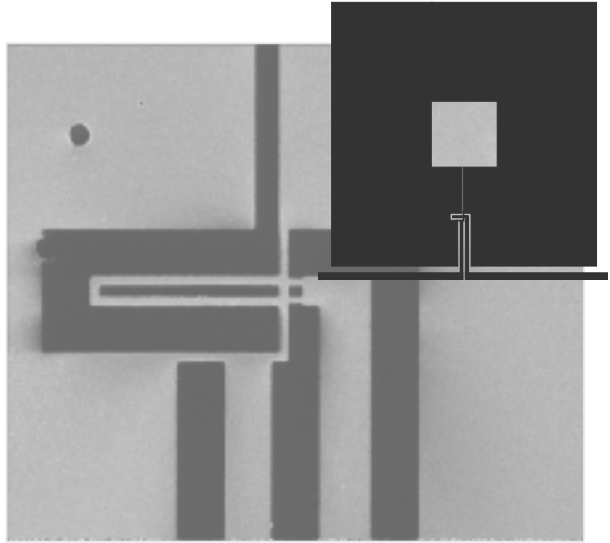

Figure 1. Burnell et al. Applied Physics Letters

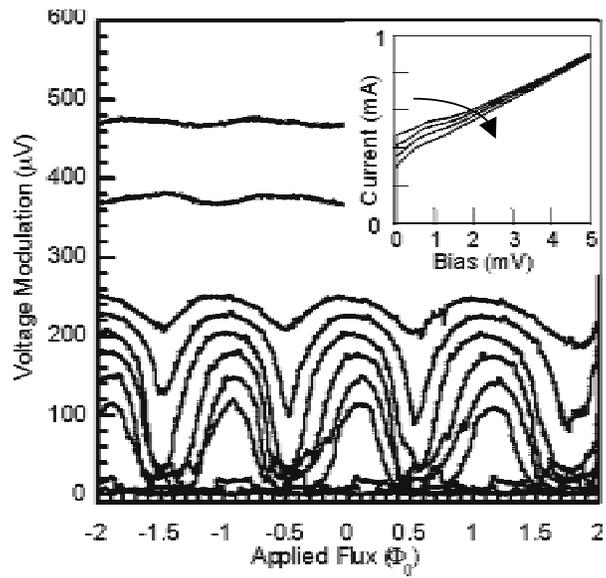

Figure 2. Burnell et al Applied Physics Letters

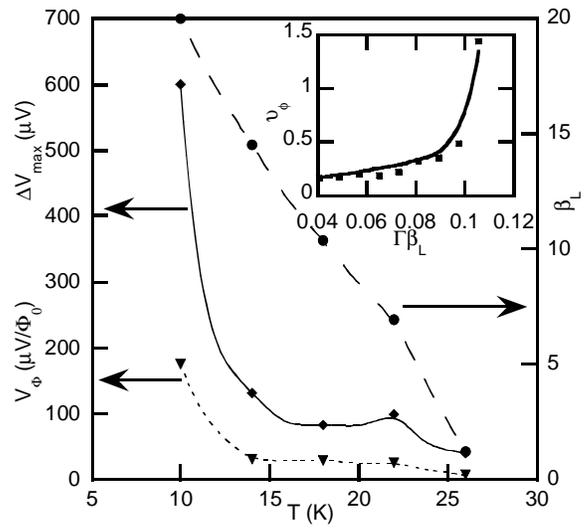

Figure 3 Burnell et al Applied Physics Letters

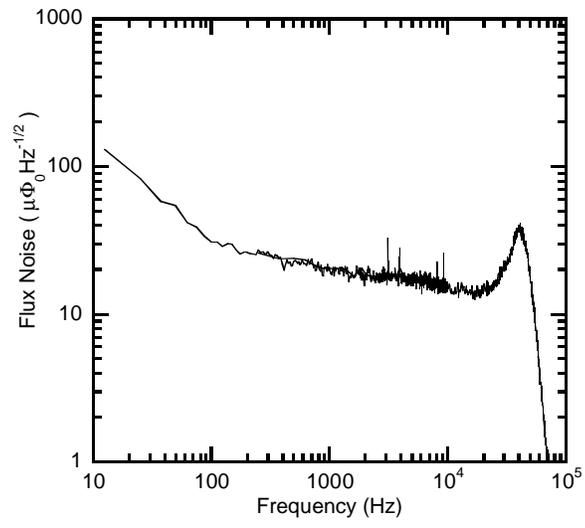

Figure 4. Burnell et al Applied Physics Letters